\documentclass[11pt,a4]{article}
\usepackage{amsmath,amsfonts,graphics,color,amssymb,amstext}
\newtheorem{thm}{Theorem}[section]
\newtheorem{lemma}[thm]{Lemma}

\newtheorem{rem}[thm]{Remark}
\newenvironment{pf} {\noindent{\sc Proof. }}{{\hfill
$\Box$}\par\vskip2\parsep}

\newcommand{\zd}{{\mathbb Z}^d}

\newcommand{\Id}{\;\text{\rm Id}}

\newcounter{mycount}

\newlength{\slimlen}%
\newcommand{\rmd}{\,\mathrm{d}}
\newcommand{\la}{\left\langle}
\newcommand{\ra}{\right\rangle}
\newcommand{\ZZ}{\ensuremath{\mathbb{Z}}}
\newcommand{\TT}{\ensuremath{\mathbb{T}}}
\newcommand{\RR}{\ensuremath{\mathbb{R}}}
\newcommand{\U}{\ensuremath{\mathbb{U}}}

\newcommand{\var}{{\mbox{\rm\normalfont var}}}
\newcommand{\cov}{{\mbox{\rm\normalfont cov\,}}}

\usepackage{graphics}

\title{Strict convexity of the free energy for non-convex gradient models at moderate $\beta$}

\author{
Codina Cotar
\thanks{Corresponding Author}
\thanks{Supported by the DFG-Forschergruppe 718 \lq Analysis and stochastics in
complex physical systems\rq}
\thanks{%
TU Berlin - Fakult\"at II,
Institut f\"ur Mathematik,
Strasse des 17. Juni 136,
D-10623 Berlin, Germany.
E-mail: {\tt cotar@math.tu-berlin.de}}
, Jean-Dominique Deuschel\footnotemark[1]\;\thanks{%
TU Berlin - Fakult\"at II
Institut f\"ur Mathematik
Strasse des 17. Juni 136
D-10623 Berlin, Germany.
E-mail: {\tt deuschel@math.tu-berlin.de}}
, and Stefan M\"uller\footnotemark[1]\;\thanks{%
Max Planck Institute for Mathematics in the Sciences
Inselstrasse 22-26
D-04103 Leipzig, Germany.
E-mail: {\tt sm@mis.mpg.de}}}

\begin{document}
\maketitle
\begin{abstract}
We consider a gradient interface model on the lattice with interaction potential which is a
non-convex perturbation of a convex potential. We show using a one-step multiple scale
analysis the strict convexity of the surface tension at high temperature. This is an extension of Funaki
and Spohn's result \cite{FS}, where the strict convexity of potential was crucial in their
proof. 
\end{abstract}

%\vspace{\fill}

         

\thispagestyle{empty}

\par

\section{ Introduction}
We consider an an effective model with gradient interaction.
The model describes a phase separation in $\Bbb R^{d+1}$, eg. between the liquid and vapor phase.
For simplicity we consider a discrete basis $\Lambda_M\subset\Bbb Z^d$, and continuous height
variables
$$x\in\Lambda_M\longrightarrow \phi(x)\in\Bbb R.$$
This model ignores overhangs like in Ising models, but gives a good approximation in the vicinity
of the phase separation. The distribution of the interface is given in terms of its Gibbs
distribution with nearest neighbor interactions of gradient type, that is,
the interaction between two neighboring sites $x,y$ depends only
on the discrete gradient,$\nabla\phi(x,y)=\phi(y)-\phi(x)$. More precisely, the Hamiltonian
is of the form
\begin{equation}
\label{tag1}
H_M(\phi)=\sum_{x,y\in \Lambda_{M+1}, |x-y|=1}V(\phi(y)-\phi(x))
\end{equation}
where $V\in C^2(\Bbb R)$ is a function with quadratic growth at infinity:
\begin{equation}
\label{tag2}
V(\eta)\ge A|\eta|^2-B,\qquad \eta\in\Bbb R
\end{equation}
for some $A>0, B\in\Bbb R$. 

For a given boundary condition $\psi\in\Bbb R^{\partial \Lambda_M}$, where
$\partial \Lambda_M=\Lambda_{M+1}\setminus \Lambda_M$,  the (finite) Gibbs distribution
on $\Bbb R^{\Lambda_{M+1}}$ at inverse temperature $\beta>0$ is given by
$$\mu_{V_{M},\psi}^{\beta}(d\phi)\equiv\frac{1}{Z_{M,\psi}^{\beta}}\exp(-\beta H_M(\phi))\prod_{x\in \Lambda_M}d\phi(x)\prod_{x\in\partial \Lambda_M}\delta_{\psi(x)}(d\phi(x)).$$
Here $Z_{M,\psi}^{\beta}$ is a normalizing constant given by
$$Z_{M,\psi}^{\beta}=\int_{\Bbb R^{\Lambda_{M+1}}}\exp(-\beta H_M(\phi))\prod_{x\in \Lambda_M}d\phi(x)
\prod_{x\in\partial \Lambda_M}\delta_{\psi(x)}(d\phi(x)).$$
One is particularly interested in tilted boundary conditions 
$$\psi_u(x)=<x,u>=\sum_{i=1}^d x_i u_i$$
for some given 'tilt' $u\in\Bbb R^d$. This corresponds to an interface in $\Bbb R^{d+1}$
which stays normal to the vector $n_u=(u,-1)\in\Bbb R^{d+1}$.

An object of basic relevance in this context is the surface tension or free energy defined by the limit
\begin{equation}
\label{tag3}
\sigma(u)=\lim_{M\to\infty}-\frac{1}{\beta}\log Z_{M,\psi_u}^{\beta}.
\end{equation}

The existence of the above limit follows from a standard sub-additivity argument. 
In fact the surface  tension 
can also be defined in terms of the partition function on the torus, see below and \cite{FS}.
In case of {\it strictly} convex potential $V$ with
\begin{equation}
\label{tag4}
c_1\le V^{''}\le c_2
\end{equation}
where $0<c_1\le c_2<\infty$, Funaki and Spohn showed in \cite{FS}
 that $\sigma$ is strictly convex.

The simplest strictly convex potential is the quadratic one with $V(\eta)=|\eta|^2$,
which corresponds to a Gaussian model, also called gradient free field or harmonic crystal.
Models with non quadratic potentials $V$ are sometimes called anharmonic crystals.

The strict convexity of the surface tension $\sigma$ plays a crucial role in the derivation of the hydrodynamical limit
 of the Landau-Ginsburg model in \cite{FS}.

 Under the condition (\ref{tag4}), a large deviation principle for the rescaled
profile with rate function given in terms of the integrated surface tension has been derived in \cite{DGI}.
Here also the strict convexity of $\sigma$ is very important.
Both papers \cite{FS} and \cite{DGI} use very explicitely the condition (\ref{tag4}) in their proof.
In particular they rely on the Brascamp Lieb
inequality and on the random walk representation of Helffer and S\"jostrand, which requires a strictly convex potential $V$.

The objective of our work is to prove strict convexity of $\sigma$ also for some non convex potential.
One cannot expect strict convexity for any non convex $V$, see below.
Our result is perturbative at high temperature (small $\beta$), and shows strict convexity of $\sigma(u)$
at every $u\in\Bbb R$
for potentials $V$ of the form
$$V(\eta)=V_0(\eta)+g_0(\eta)$$
where $V_0$ satisfies (\ref{tag4}) and $g_0\in C^2(\Bbb R)$ has a negative bounded second derivative
such that $\sqrt{\beta}\cdot \|g_0^{''}\|_{L^1(\RR)}$ is small enough.

Our proof is based on the scale decomposition of the free field as the sum of two
independent free fields $\phi_1$ and $\phi_2$, where we choose the variance of $\phi_1$
small enough to match the non-convexity of $g$. This particular type of scale decomposition was used earlier by Haru Pinson in \cite{Pins}, who also 
suggested to us the use of this approach.
%JDD
The partition function $Z_{N,\psi_u}^{\beta}$ can be then expressed
in terms of a double integral, with respect to both $\phi_1$ and $\phi_2$. We fix $\phi_2$ and perform
first the integration with respect to $\phi_1$. This yields a new induced Hamiltonian, which is a function of the remaining variable
$\phi_2$.
The main point is that our choice of the variance
of $\phi_1$ and smallness of $\beta$  allow us to show convexity in $\phi_2$ of the induced Hamiltonian. Of course this Hamiltonian
is no longer of the simple form (\ref{tag1}), in particular we lose the locality of the interaction.
However an extension of the technique introduced in \cite{DGI} shows strict convexity of $\sigma.$
The idea behind the proof is that one can gain convexity via integration.
This procedure is called "one step decomposition", since we perform only one integration.
Of course this procedure could be iterated which would allow to lower the temperature.
However for general non convex $g$ we do not expect that this procedure works at low temperature for every tilt $u$.

At low temperature an approach in the spirit of \cite{BY}, \cite {B} looks more promising \cite{AKM}. 

Finally note that, due to the gradient interaction, the Hamiltonian has a continuous symmetry. In particular
this implies that no infinite Gibbs state exists for the lower lattice dimensions, $d=1,2$
where the field "delocalizes" as $M\to\infty$,
%JDD
 c.f. \cite{FP}.
On the other hand, it is very natural in this setting to consider the {\it gradient} Gibbs distributions,
that is the image of 
%JDD
$\mu_{V_{M},\psi}$ under the gradient operation $\phi\in\Bbb R^{\Bbb Z^d}\longrightarrow
\nabla\phi$. It is easy to verify that this distribution depends only on $\nabla\psi$, the gradient of the 
boundary condition, in fact one can also introduce gradient Gibbs distributions in terms of conditional 
distributions satisfying DLR conditions, c.f. \cite{FS}. Using the quadratic bound (\ref{tag1}), one can  easily
see that the corresponding measures are tight. In particular for each tilt $u\in\Bbb R^d$ one can construct
a translation invariant gradient Gibbs state $\tilde\mu_u$ on $\Bbb Z^d$ with mean $u$:
 $$\Bbb E_{\tilde\mu_u}[\phi(y)-\phi(x)]=<y-x,u>.$$

Under (\ref{tag4}), Funaki and Spohn proved the existence and unicity of extremal, ie. ergodic, gradient Gibbs state, for each tilt $u\in\Bbb R$. 
In the case of non convex $V$, unicity of the ergodic states can be violated, even at $u=0$ tilt, c.f. \cite{BK}.
However in this situation, the surface tension is not strictly convex at $u=0$.

\section{Main result and outline of the proof}

We study the convexity properties of the free energy (as a function of the tilt $u$) for non-convex gradient models on a lattice.
%JDD why torus
 Using the results of \cite{FS}, we work on the torus, instead of the box $\Lambda_M$, see Remark 2.4 below.
Thus, let $\TT_M^d=(\ZZ/M\ZZ)^d=\zd~\mbox{mod}~(M)$ be the lattice torus in $\zd$, let $u\in\RR^d$ and let $\beta>0$. For a function $\phi:\TT_M^d\rightarrow\RR$, we consider the discrete derivative
\begin{eqnarray}
\nabla_i\phi(x)=\phi(x+e_i)-\phi(x)
\end{eqnarray}
and the Hamiltonian
%JDD u missing!
%
\begin{eqnarray}
H(u,\phi)=\sum_{x\in\TT_M^d}\sum_{i=1}^d\left[V_0^i(\nabla_i\phi(x)+u_i)+g_0^i(\nabla_i\phi(x)+u_i)\right],
\end{eqnarray}
where $V_0^i$ is convex and $g_0^i$ is non-convex (see (\ref{gVcond}) below). We consider the partition function
\begin{eqnarray}
Z_M^{\beta}(u)=\int_X e^{-\beta H(u,\phi)}m_M(\rmd\phi),
\end{eqnarray}
where
\begin{eqnarray}
X=\{\phi:\TT_M^d\rightarrow\RR:\phi(0)=0\}
\end{eqnarray}
and
\begin{eqnarray}
m_M(\rmd\phi)=\prod_{x\in\TT_M^d\setminus \{0\}}\rmd\phi(x)\delta_0(\rmd\phi(0)),
\end{eqnarray}
and the free energy
\begin{eqnarray}
\label{surfte}
f_M^{\beta}(u)=-\frac{1}{\beta}\log Z_M^{\beta}(u).
\end{eqnarray}
We will prove
\begin{thm}
\label{main}
Suppose that $V_0^i$ and $g_0^i$ are $C^2$ functions on $\RR$ and that there exist constants $C_0,C_1,C_2$ and
\begin{eqnarray}
\label{gVcond}
0<C_1\le (V_0^i)''\le C_2,~~-C_0\le (g_0^i)''\le 0.
\end{eqnarray}
Set
\begin{eqnarray}
\label{barc}
\bar{C}=\max\left(\frac{C_0}{C_1},\frac{C_2}{C_1}-1,1\right).
\end{eqnarray}
If $(g_0^i)''\in L^1(\RR)$ and for $i\in\{1,2,\ldots,d\}$
\begin{eqnarray}
\label{fcond}
\frac{4}{\pi}(12d\bar{C})^{1/2}\sqrt{\beta C_1}\frac{1}{C_1}||(g_0^i)''||_{L^1(\RR)}\le\frac{1}{2},
\end{eqnarray}
then
\begin{eqnarray}
\label{hessianf}
(D^2f_M^{\beta})(u)\ge\frac{C_1}{2}|\TT_M^d| \Id,~\forall u\in\RR^d,
\end{eqnarray} 
where $|\TT_M^d|=M^d$ denotes the number of points in $\TT_M^d$. In other words, the free energy per particle is uniformly convex, uniformly in $M$.
\end{thm}
\begin{rem}\normalfont
The main point is that the convexity estimate (\ref{hessianf}) holds uniformly in the size $M$ of the torus. Indeed a direct calculation of $D^2f_M^1$ yields at $u$
\begin{eqnarray}
D^2f_M^1(u)=\la D^2_u H(u,\cdot)\ra_H-\var_H D_uH (u,\cdot),
\end{eqnarray}
where
%JDD u missing
\begin{eqnarray}
\la f\ra_H=\frac{\int_X f(\phi)e^{-H(u,\phi)}m_M(\rmd\phi)}{\int_X e^{-H(u,\phi)}m_M(\rmd\phi)}
\end{eqnarray}
and
\begin{eqnarray}
\var_H f=\la \left(f-\la f\ra_H\right)^2\ra_H.
\end{eqnarray}
Now one might expect that a condition like (\ref{fcond}) implies that $\la (D^2_u H(u,\cdot)\ra_H\ge cC_1 |\TT_M^d|\Id$ (see Lemma~\ref{l1norm} below). The problem 
is that naively the variance term scales like $|\TT_M^d|^2$ since $D_uH$ is a sum of $d|\TT_M^d|$ terms. To get a better estimate, one has to show that in a suitable sense, the terms
\begin{eqnarray}
\cov_H\left(D_u(V_0+g_0)(u+\nabla_i\phi(x)),D_u(V_0+g_0)(u+\nabla_j\phi(y))\right)
\end{eqnarray}
decay if $|x-y|$ is large. If $H$ is not convex such a decay of correlations is, presently, only proved for the class of potentials studied in \cite{cojd}. 
As discussed above, the Helffer-Sj\"ostrand estimates do not apply directly. The main idea is to rewrite $Z_M^{\beta}(u)$ as an iterated integral in such a way that each integration involves a convex hamiltonian to which the Helffer-Sj\"ostrand theory can be applied (see (\ref{19}) below).
\end{rem}
\begin{rem}\normalfont
Instead of $||g_0''||_{L^1(\RR)}$ one can also use bounds on lower order derivatives. More precisely, condition (\ref{fcond}) can, for example be replaced by
\begin{eqnarray}
\label{9}
\frac{50}{\sqrt{2\pi}}d\bar{C}(\beta C_1)^{3/4}\frac{1}{C_1}||g_0'||_{L^2(\RR)}\le\frac{1}{2}
\end{eqnarray}
(see Remark~\ref{3.2} below). In view of the estimate
\begin{eqnarray}
\int_{\RR} (g_0')^2(s)\rmd s=\int_{\RR} g_0(s)g_0''(s)\rmd s\le C_0||g_0||_{L^1(\RR)},
\end{eqnarray}
we can see that (\ref{fcond}) can be replaced by
\begin{eqnarray}
cd^2\bar{C}^3 (\beta C_1)^{3/2}\frac{1}{C_1}||g_0||_{L^1(\RR)}\le\frac{1}{4}
\end{eqnarray}
with $c=\frac{2500}{2\pi}$.
\end{rem}
\begin{rem}\normalfont
Note that the surface tensions defined in (\ref{tag3}) and (\ref{surfte}) coincide (see, for example, \cite{FS}). Because of this, we will work from now on with the definition of the surface tension on a torus, as it is easier to use.
\end{rem}

\subsection*{Outline of the proof for Theorem~\ref{main}}

\subsubsection*{Step 1: Scaling argument}

A simple scaling argument shows that it suffices to prove the result for 
\begin{eqnarray}
\label{11}
\beta=1, C_1=1.
\end{eqnarray}
Indeed, suppose that the result is true for $\beta=1$ and $C_1=1$. Given $\beta$, $V_0^i$ and $g_0^i$ which satisfy (\ref{gVcond}) and (\ref{fcond}), we define
\begin{eqnarray}
\tilde{V_0^i}(s)=\beta V_0^i\left(\frac{s}{\sqrt{\beta C_1}}\right),~\tilde{g_0^i}(s)=\beta g_0^i\left(\frac{s}{\sqrt{\beta C_1}}\right).
\end{eqnarray}
Then
$$1\le (\tilde{V_0^i})''\le\frac{C_2}{C_1},~-\frac{C_0}{C_1}\le (\tilde{g_0^i})''\le 0,$$
\begin{equation}
||(\tilde{g_0^i})''||_{L^1(\RR)}=\sqrt{\beta C_1}\frac{1}{C_1}||(g_0^i)''||_{L^1(\RR)}.
\end{equation}
Hence $\tilde{V_0^i}$, $\tilde{g_0^i}$ satisfy the assumptions of Theorem~\ref{main} with $\beta=1$ and $C_1=1$. Thus
\begin{eqnarray}
\label{12}
D^2f_M^1(\cdot,\tilde{V_0^i},\tilde{g_0^1})\ge\frac{1}{2}|\TT_M^d| \Id.
\end{eqnarray}
On the other hand, the change of variables
\begin{eqnarray}
\tilde{\phi}(x)=\sqrt{\beta C_1}\phi(x),\tilde{u}=\sqrt{\beta C_1} u
\end{eqnarray}
yields
\begin{eqnarray}
\tilde{V_0^i}\left(\tilde{u_i}+\nabla_i\tilde{\phi}(x)\right)=V_0^i(u+\nabla_i\phi(x))
\end{eqnarray}
and thus
\begin{eqnarray}
Z_M^{\beta}(u,V_0^i,g_0^i)=(\beta C_1)^{-(|\TT_M^d|-1)/2}Z_M^1(\tilde{u},\tilde{V_0},\tilde{g_0}).
\end{eqnarray}
Hence
\begin{eqnarray}
f_M^{\beta}(u,V_0,g_0)=const(\beta,C_1)+\frac{1}{\beta}f^1_M\left(\sqrt{\beta C_1} u, \tilde{V_0},\tilde{g_0}\right).
\end{eqnarray}
Thus (\ref{12}) implies (\ref{fcond}), as claimed.

\subsubsection*{Step 2: Separation of the Gaussian part}

Next we separate the Gaussian part in the Hamiltonian. From now on, we will always assume that $\beta=1$ and $C_1=1$. Set
\begin{eqnarray}
\label{13}
V_1(s)=V_0(s)-\frac{1}{2}s^2,~g=V_1+g_0.
\end{eqnarray}
Then
\begin{eqnarray}
\label{14}
0\le V_1''\le C_2-1,~~-C_0\le g''\le C_2-1
\end{eqnarray}
and the Hamiltonian can be rewritten as
\begin{eqnarray}
\label{15}
H(u,\phi)=\sum_{x\in\TT_M^d}\sum_{i=1}^d\frac{1}{2}\left(u_i+\nabla_i\phi(x)\right)^2+G(u,\phi),
\end{eqnarray}
where
\begin{eqnarray}
G(u,\phi)=\sum_{x\in\TT_M^d}\sum_{i=1}^d g(u_i+\nabla_i\phi(x)).
\end{eqnarray}
Since for all functions $\phi$ on the torus and for all $i\in\{1,2,\ldots d\}$
\begin{eqnarray}
\sum_{x\in\TT_M^d}\nabla_i\phi(x)=0,
\end{eqnarray}
we get
\begin{eqnarray}
H(u,\phi)=\frac{1}{2}|\TT_M^d||u|^2+\frac{1}{2}||\nabla\phi||^2+G(u,\phi),
\end{eqnarray}
where $||\nabla\phi||^2=\sum_{x\in\TT_M}\sum_{i=1}^d |\nabla_i\phi(x)|^2$. Let
\begin{eqnarray}
Z_0=\int_X e^{-\frac{1}{2}||\nabla\phi||^2}m_M(\rmd\phi).
\end{eqnarray}
Then the measure
\begin{eqnarray}
\label{16}
\mu=\frac{1}{Z_0}e^{-\frac{1}{2}||\nabla\phi||^2}m_M(\rmd\phi)
\end{eqnarray}
is a Gaussian measure. Its covariance $C$ is a positive definite symmetric operator on $X$ (equipped with a standard scalar product $(\phi,\psi)=\sum_{x\in\TT_M^d}\phi(x)\psi(x)$) such that
\begin{eqnarray}
\label{17}
(C^{-1}\phi,\phi)=||\nabla\phi||^2,~\forall\phi\in X.
\end{eqnarray}
The partition function thus becomes (recall that we take $\beta=1$)
\begin{eqnarray}
\label{18}
Z_M(u)=Z_0 e^{-\frac{1}{2}|\TT_M^d||u|^2}\int_X e^{-G(u,\phi)}\mu(\rmd\phi).
\end{eqnarray}

\subsubsection*{Step 3: Decomposition of $\mu$ and Helffer-Sj\"ostrand calculus}

By standard Gaussian calculus, $\mu=\mu_1\ast\mu_2$, where $\mu_1$ and $\mu_2$ are Gaussian with covariances
\begin{eqnarray}
C_1=\lambda C,~C_2=(1-\lambda)C,~~\mbox{where}~~\lambda\in (0,1).
\end{eqnarray}
More explicitly, for $i\in\{1,2\}$
\begin{eqnarray}
\mu_i(\rmd\phi)=\frac{1}{Z_i}e^{-\frac{1}{2\lambda_i}||\nabla\phi||^2}m_M(\rmd\phi),~~\mbox{where}~~\lambda_1=\lambda,\lambda_2=1-\lambda.
\end{eqnarray}
Thus
\begin{eqnarray}
\label{19}
Z_M(u)=Z_0e^{-\frac{1}{2}|\TT_M^d||u|^2}\int_X\int_X e^{-G(u,\psi+\theta)}\mu_1(\rmd\theta)\mu_2(\rmd\psi).
\end{eqnarray}
To write the free energy in a more compact form, we introduce the renormalization maps $R_i$. For $f\in C(\RR^d\times X)$ we define $R_i f$ by
\begin{eqnarray}
e^{-R_if(u,a)}:=\int_X e^{-f(u,a+b)}\rmd\mu_i(b).
\end{eqnarray}
Taking the logarithm of (\ref{19}), we get
\begin{eqnarray}
\label{20}
f_M(u)=const(M)+\frac{1}{2}|\TT_M^d||u|^2+(R_2R_1G) (0,u).
\end{eqnarray}
The main point now is that the map
\begin{eqnarray}
H_1(\theta)=G(u,\psi+\theta)+\frac{1}{2\lambda}||\nabla\theta||^2
\end{eqnarray}
becomes uniformly convex for sufficiently small $\lambda$. This will allow us to use the Helffer-Sj\"ostrand representation to get a good lower bound for 
$D^2(R_1G)$, which involves, roughly speaking, the expectation of $G_{i,x}(\theta)=g_0''(u_i+\nabla_i\psi(x)+\nabla_i\theta(x))$ with respect to 
$e^{-H_1}$ (see (\ref{c4})). This expectation can be controlled in terms of $||g_0''||_{L^1(\RR)}$ (see Lemma~\ref{l1norm}). Under the smallness condition (\ref{fcond}) one then easily obtains the lower bound for $D^2(R_2R_1G)$ (see (\ref{c6}) and (\ref{c7})).

\section{Consequence of the Helffer-Sj\"ostrand representation}

Let $\U$ and $X$ be finite-dimensional inner product spaces, let $C$ be a positive definite symmetric operator on $X$ and let $\mu_C$ be the Gaussian 
measure with covariance $C$ on $X$, i.e
\begin{eqnarray}
\mu_C(db)=\frac{1}{Z_C}e^{-\frac{1}{2}(C^{-1}b,b)}\rmd b,
\end{eqnarray}
where $\rmd b$ is the dim X dimensional Hausdorff measure on $X$ (i.e $\rmd b=\prod\rmd b_i$ if the $b_i$ are the coordinates with respect to an orthonormal 
basis). For a continuous function $f\in C(\U\times X)$ we define $R_{C}f$ by
\begin{eqnarray}
e^{-R_{C}f(u,a)}=\int_X e^{-f(u,a+b)}\rmd\mu_{C}(\rmd b).
\end{eqnarray}
In the situation we will consider, $b\rightarrow f(u,a+b)+\frac{1}{2}(C^{-1}b,b)$ will be convex and hence bounded from below so that the right hand side 
of the above identity is strictly positive.

For $f\in C^2(\U\times X)$ we write $D^2 f(u,a)$ for the Hessian at $(u,a)$, viewed as an operator from $\U\times X$ to itself. The restriction of the Hessian to $X$ is denoted by $D^2_Xf:=P_X D^2f P_X$, where $P_X$ is the orthogonal projection $\U\times X\rightarrow X$. On the level of quadratic forms we thus have
\begin{eqnarray}
\left(D^2_X f(u,a)(\dot{u},\dot{a}),(\dot{u},\dot{a})\right)=\left(D^2f(u,a)(0,\dot{a}),(0,\dot{a})\right).
\end{eqnarray}
From the Helffer-Sj\"ostrand representation of the variance (see, e.g., \cite{He} (2.6.15)) and the duality relation
\begin{eqnarray}
\frac{1}{2}\left (A^{-1}a,a\right)=\sup_{b\in D(A^{\frac{1}{2}})}\left((a,b)-\frac{1}{2}(Ab,b)\right),
\end{eqnarray}
which holds for any positive definite self-adjoint operator $A$ on a Hilbert space $Y_0$, one immediately obtains the following estimate:
\begin{lemma} 
\label{HS/DGI}
Supppose that $H\in C^2(X)$, $\sup_X|D^2H|<\infty$ and there exists a $\delta>0$ such that
\begin{eqnarray}
D^2H(a)\ge\delta \Id,~~\forall a\in X.
\end{eqnarray}
Set
\begin{eqnarray}
Y_0=\{K\in L^2_{loc}(X):\la|DK|^2\ra_H<\infty\},
\end{eqnarray}
%JDD
\begin{eqnarray}
\label{b2}
Y=\{K\in Y_0:\la||D^2K||^2_{HS}\ra_H<\infty\},
\end{eqnarray}
where the derivatives are understood in the weak sense and
%JDD
\begin{eqnarray}
\|DK^2\|^2_{HS}:=\sum_{x,y\in\TT_M^d\setminus\{0\}}\left(\frac{\partial^2}{\partial\phi(x)\partial\phi(y)} K\right)^2
\end{eqnarray}
denotes the Hilbert-Schmidt norm. Then for all $G\in Y$ we have
\begin{eqnarray}
\var_H G=\sup_{K\in Y}\la 2(DG,DK)-(DK,D^2H DK)-\|D^2 K\|^2_{HS}\ra_H
\end{eqnarray}
Therefore
\begin{eqnarray}
\label{dgi}
\var_H G\le\sup_{K\in Y}\la 2(DG,DK)-(DK,D^2H DK)\ra_H.
\end{eqnarray}
\end{lemma}
We will use (\ref{dgi}) from Lemma~\ref{HS/DGI} in the proof of the lemma below.
\begin{lemma}
\label{hessian}
Suppose that $f\in C^2(\U\times X)$ and $\sup_{\U\times X}|D^2 f|<\infty$. Suppose moreover that there exists a $\delta>0$ such that
\begin{eqnarray}
\label{b1}
D^2f(u,a)+C^{-1}\ge\delta \Id, ~~\forall (u,a)\in\U\times X.
\end{eqnarray}
Then $Rf\in C^2(\U\times X)$ and for all $u,\dot{u}\in\U,a,\dot{a}\in X$
\begin{eqnarray}
\label{b3}
\lefteqn{\left((D^2Rf)(u,a)(\dot{u},\dot{a}),(\dot{u},\dot{a})\right)}\nonumber\\
&\ge&\inf_{K\in Y}\la\left(D^2 f(u,a+\cdot)(\dot{u},\dot{a}-DK(\cdot),(\dot{u},\dot{a}-DK(\cdot))\right) \ra_{H,a}\nonumber\\
&+&\la (C^{-1}DK(\cdot),DK(\cdot))\ra_{H_{u,a}}
\end{eqnarray}
where
\begin{eqnarray}
H_{u,a}(b)=f(u,a+b)+\frac{1}{2}(C^{-1}b,b),
\end{eqnarray}
\begin{eqnarray}
\la g\ra_{H_{u,a}}=\frac{\int g(b) e^{-H_{u,a}}(b)\rmd b}{\int e^{-H_{u,a}(b)}\rmd b}.
\end{eqnarray}
\end{lemma}
{\em Proof}

We have
\begin{eqnarray}
\label{b4}
e^{-Rf(u,a)}=\int_X e^{-\left[f(u,a+b)+\frac{1}{2}(C^{-1}b,b)\right]}\rmd b.
\end{eqnarray}
It follows from (\ref{b1}) that
\begin{eqnarray}
f(u,a+b)+\left(C^{-1}(a+b),(a+b)\right)\ge\frac{1}{2}\delta|a+b|^2-c
\end{eqnarray}
and standard estimates yield
\begin{eqnarray}
f(u,a+b)+(C^{-1}b,b)\ge\frac{1}{4}\delta|b|^2-c\left(1+|a|^2\right).
\end{eqnarray}
Hence, by the dominated convergence theorem, the right-hand side of (\ref{b4}) is a $C^2$ function in $(u,a)$ and the same applies to $Rf$ since the 
right-hand side of (\ref{b4}) does not vanish.

~

To prove the estimate (\ref{b3}) for $D^2Rf$, we may assume without loss of generality that $a=0,u=0$ (otherwise we can consider the shifted function $f(\cdot-u,\cdot-a)$). Set
\begin{eqnarray}
h(t):=Rf(t\dot{u},t\dot{a}).
\end{eqnarray}
Then
\begin{eqnarray}
h''(0)=\left(D^2(Rf)(0,0)(\dot{u},\dot{a}),(\dot{u},\dot{a})\right).
\end{eqnarray}
Now 
\begin{eqnarray}
h(t)=-\log\int_X e^{-f(t\dot{u},t\dot{a}+b)}\mu_C(\rmd b),
\end{eqnarray}
\begin{eqnarray}
h'(t)=\frac{\int_Xe^{-f(t\dot{u},t\dot{a}+b)}Df(t\dot{u},t\dot{a}+b)(\dot{u},\dot{a})\mu_C(\rmd b)}{\int_Xe^{-f(t\dot{u},t\dot{a}+b)}\mu_C(\rmd b)}
\end{eqnarray}
and
\begin{eqnarray}
\label{b5}
h''(0)=\la\left(D^2 f(0,\cdot)(\dot{u},\dot{a}),(\dot{u},\dot{a})\right)\ra_H-\var_H Df(0,\cdot)(\dot{u},\dot{a}),
\end{eqnarray}
where
\begin{eqnarray}
H(b)=f(0,b)+\frac{1}{2}(C^{-1}b,b).
\end{eqnarray}
By assumption,
\begin{eqnarray}
D^2H(b)\ge\delta \Id,
\end{eqnarray}
i.e. $H$ is uniformly convex.
\endpf
Hence by (\ref{dgi}) from Lemma~\ref{HS/DGI}
\begin{eqnarray}
\label{b6}
-\var_Hg\ge\inf_{K\in Y}\la-2(Dg,DK)+(DK,D^2HDK)\ra_H.
\end{eqnarray}
Apply this with
\begin{eqnarray}
g(b)=Df(0,b)(\dot{u},\dot{a})
\end{eqnarray}
and write
\begin{eqnarray}
D^2H=D^2_Xf+C^{-1}.
\end{eqnarray}
Then
\begin{eqnarray}
\lefteqn{-2(Dg,DK)+(DK,D^2HDK)}\nonumber\\
&=&-2D^2f(0,\cdot)\left((\dot{u},\dot{a}),(0,DK)\right)+D^2f(0,\cdot)\left((0,DK),(0,DK)\right)\nonumber\\
&&+(C^{-1}DK,DK).
\end{eqnarray}
Together with (\ref{b6}) and (\ref{b5}) this yields (\ref{b3}).
\endpf

\section{Proof of Theorem~\ref{main}}
By (\ref{20})
\begin{eqnarray}
\label{c1}
f_M(u)=const(M)+\frac{1}{2}|\TT_M^d||u|^2+(R_2R_1G)(0,u),
\end{eqnarray}
where
\begin{eqnarray}
G(u,\phi)=\sum_{x\in\TT_M^d}\sum_{i=1}^dg^i(u_i+\nabla_i\phi).
\end{eqnarray}
We first estimate $D^2R_1G$ from below. By (\ref{14})
\begin{eqnarray}
\label{c2}
(g^i)''\ge -C_0\ge-\bar{C}
\end{eqnarray}
(recall that we always assume $C_1=1)$. By (\ref{barc}), we have $\bar{C}\ge 1$. If we take
\begin{eqnarray}
\label{c3}
\lambda=\frac{1}{2\bar{C}}
\end{eqnarray}
then
\begin{eqnarray}
H_{u,\psi}(\theta):=G(u,\psi+\theta)+\frac{1}{\lambda}||\nabla\theta||^2
\end{eqnarray}
is uniformly convex, i.e.
\begin{eqnarray}
D^2H_{u,\psi}(\theta)(\dot{\theta},\dot{\theta})\ge\bar{C}||\nabla\dot{\theta}||^2\ge\delta_M\bar{C}||\dot{\theta}||^2,
\end{eqnarray}
with $\delta_M>0$. Here we used the discrete Poincare inequality 
\begin{eqnarray}
||\nabla\eta||^2\ge\delta_M||\eta||^2	~~\mbox{for}~~\eta\in X
\end{eqnarray}
which follows from a simple compactness argument since $\TT_M^d$ is a finite set.
Hence, by Lemma~\ref{hessian}, we have
\begin{eqnarray}
\lefteqn{\left(D^2R_1(G)(u,\psi)(\bar{u},\bar{\psi}),(\bar{u},\bar{\psi})\right)}\nonumber\\
&\ge&\inf_{K\in Y}\bigg\{\la\sum_{x\in\TT_M^d}\sum_{i=1}^d (g^i)''\left(u_i+\nabla_i\psi(x)+\nabla_i\cdot(x)\right)\right.
\left( u_i+\nabla_i\psi(x)-\nabla_i\frac{\partial K}{\partial\phi(x)}(\cdot)\right)^2\nonumber\\
&&\left.+\frac{1}{\lambda}\sum_{x\in\TT_M^d}\sum_{i=1}^d\left|\nabla_i\frac{\partial K}{\partial\phi(x)}\right|^2\ra_{H_{u,\psi}}\bigg\},
\end{eqnarray}
where $Y$ is defined by (\ref{b2}).
Now $(g^i)''=(V^i)''+g_0''\ge g_0''$ (see (\ref{13}) and (\ref{14})) and together with the estimate $(a-b)^2\le 2a^2+2b^2$ and the assumption 
$-C_0\le g_0''\le 0$, this yields
\begin{eqnarray}
\label{c4}
\lefteqn{\left(D^2R_1(G)(u,\psi),(\dot{u},\dot{\psi}),(\dot{u},\dot{\psi})\right)}\nonumber\\
&\ge&2\sum_{x\in\TT_m^d}\sum_{i=1}^d\la (g_0^i)''(u_i+\nabla_i\psi(x)+\nabla_i\cdot(x))\left(u_i+\nabla_i\psi(x)\right)^2\ra_{H_{u,\psi}}\nonumber\\
&&+\la\left(\frac{1}{\lambda}-2C_0\right)\sum_{x\in\TT_M^d}\sum_{i=1}^d\left|\nabla_i\frac{\partial K}{\partial\phi(x)}(\cdot)\right|^2\ra_{H_{u,\psi}},
\end{eqnarray}
where $\frac{1}{\lambda}-2C_0\ge 0$. We will now use the following result, which will be proven at the end of this section.
\begin{lemma}
\label{l1norm}
For $h\in L^1(\RR)\cap C^0(\RR)$, $\psi\in X$, $x\in\TT_M^d$ and $i\in\{1,2,\ldots d\}$ consider $F\in C(X)$ given by
\begin{eqnarray}
F(\theta)=h(u_i+\nabla_i\psi(x)+\nabla_i\theta(x)).
\end{eqnarray}
Then
\begin{eqnarray}
\label{5}
\left|\la F\ra_{H_{u,\psi}}\right|\le\frac{2}{\pi}(12d\bar{C})^{1/2}||h||_{L^1(\RR)}.
\end{eqnarray}
\end{lemma}
Together with (\ref{c4}), the smallness condition (\ref{fcond}) and the relation $\sum_{x\in\TT_M^d}\nabla_i\psi(x)=0$, this lemma yields
\begin{eqnarray}
\label{c6}
\lefteqn{D^2R_1G(u,\psi)(\dot{u},\dot{\psi})(\dot{u},\dot{\psi})}\nonumber\\
&\ge& -\frac{1}{2}\sum_{x\in\TT_M^d}\sum_{i=1}^d\left|\dot{u}_i+\nabla_i\dot{\psi}(x)\right|^2=-\frac{1}{2}|\TT_M^d||\dot{u}|^2-\frac{1}{2}||\nabla\dot{\psi}||^2.
\end{eqnarray}
Thus 
\begin{eqnarray}
H_2(\psi):=(R_1G)(u,\psi)+\frac{1}{2(1-\lambda)}||\nabla\psi||^2
\end{eqnarray}
is uniformly convex and another application of Lemma~\ref{hessian} gives
\begin{eqnarray}
\label{c7}
\lefteqn{\left(D^2(R_2R_1G)(u,0)(\dot{u},0),(\dot{u},0)\right)}\nonumber\\
&\ge&\inf_K\la D^2(R_1G)(u,\cdot)(\dot{u},-DK)(\dot{u},-DK)+\frac{1}{1-\lambda}||\nabla DK||^2\ra_{H_2}\nonumber\\
&\ge&-\frac{1}{2}|\TT_M^d||\dot{u}|^2+\inf_K\bigg\{\left(\frac{1}{1-\lambda}-\frac{1}{2}\right)\la||\nabla DK||^2\ra_{H_2}\bigg\}\nonumber\\
&\ge&-\frac{1}{2}|\TT_m^d||\dot{u}|^2,
\end{eqnarray}
where in the last inequality we used that fact that $\frac{1}{1-\lambda}-\frac{1}{2}\ge 0$. In view of (\ref{c1}), this finishes the proof of 
Theorem~\ref{main}.
\subsubsection*{Proof of Lemma~\ref{l1norm}}
Note that $u$ and $\psi$ are fixed. Since the function $\tilde{h}(s)=h(u_i+\nabla_i\psi(x)+s)$ has the same $L^1$ norm as $h$, it suffices to prove the estimate for the function $F\in C(X)$ given by
\begin{eqnarray}
F(\theta)=h(\nabla_i\theta(x)).
\end{eqnarray}
Moreover, we write $H$ instead of $H_{u,\psi}$. Let
\begin{eqnarray}
\hat{h}(k)=\int_{\RR}e^{-iks}h(s)\rmd s
\end{eqnarray}
denote the Fourier transform of $h$. Then
\begin{eqnarray}
\label{c8}
||\hat{h}||_{L^{\infty}(\RR)}\le ||h||_{L^1(\RR)}
\end{eqnarray}
and
\begin{eqnarray}
h(s)=\frac{1}{2\pi}\int_{\RR}e^{iks}\hat{h}(s)\rmd k.
\end{eqnarray}
Set
\begin{eqnarray}
A(k)=\la F_k\ra_H,~~\mbox{where}~~F_k(\theta)=e^{ik\nabla_i\theta(x)}.
\end{eqnarray}
Then
\begin{eqnarray}
\label{c9}
\la F\ra_H=\frac{1}{2\pi}\int_{\RR} A(k) h(k)\rmd k
\end{eqnarray}
and, in view of (\ref{c8}), it suffices to show that
\begin{eqnarray}
\label{c10}
\int_{\RR}|A(k)|\rmd k\le 4(12 d\bar{C})^{1/2}.
\end{eqnarray}
First note that $|F_k|=1$. Hence
\begin{eqnarray}
\label{c11}
|A(k)|\le 1,~~\forall k\in\RR.
\end{eqnarray}
To get decay of  $A(k)$ for large $k$ we use integration by parts. First note that for $G_i\in C^1(X)$, with 
%JDD
$\sup_{a\in X} e^{-\delta |a|}(|G_i|(a)+|DG_i|(a))<\infty$ for all $\delta>0$, we have
\begin{eqnarray}
\label{c12}
\la\frac{\partial G_1}{\partial\phi(x)}G_2\ra_H=\la -G_1\frac{\partial G_2}{\partial\phi(x)}\ra_H+\la\frac{\partial H}{\partial\phi(x)} G_1G_2\ra_H.
\end{eqnarray}
Assume first that $x\in\TT_M^d\setminus\{0\}$. Then 
\begin{eqnarray}
F_k(\theta)=-\frac{1}{k^2}\frac{\partial^2 F_k}{\partial\theta^2(x)}(\theta)
\end{eqnarray}
and thus
\begin{eqnarray}
-k^2 A(k)&=&\la\frac{\partial^2 F_k}{\partial\theta^2(x)}\cdot 1\ra_H=\la\frac{\partial F_k}{\partial\theta(x)}\frac{\partial H}{\partial\theta(x)}\ra_H\nonumber\\
&=&-\la F_k\frac{\partial^2H}{\partial\theta^2(x)}\ra_H+\la F_k\left(\frac{\partial H}{\partial\theta(x)}\right)^2\ra_H.
\end{eqnarray}
Since $|F_k|=1$, this yields
\begin{eqnarray}
|A(k)|\le\frac{1}{k^2}\la\left|\frac{\partial^2H}{\partial\theta^2(x)}\right|\ra_H+\frac{1}{k^2}\la\left(\frac{\partial H}{\partial\theta(x)}\right)^2\ra_H.
\end{eqnarray}
Application of (\ref{c12}) with $G_2=1, G_1=\frac{\partial H}{\partial\theta(x)}$ gives
\begin{eqnarray}
\la\frac{\partial^2H}{\partial\theta^2(x)}\ra_H=\la\left(\frac{\partial H}{\partial\theta(x)}\right)^2\ra_H.
\end{eqnarray}
Thus
\begin{eqnarray}
|A(k)|\le\frac{2}{k^2}\la\left|\frac{\partial^2 H}{\partial\theta^2(x)}\right|\ra_H.
\end{eqnarray}
Now recall that
\begin{eqnarray}
H(\theta)=\sum_{x\in\TT_M^d}\sum_{i=1}^dg^i\left(u_i+\nabla_i\psi(x)+\nabla_i\theta(x)\right)+\frac{1}{2\lambda}|\nabla_i\theta(x)|^2.
\end{eqnarray}
Since $\lambda^{-1}=2\bar{C}$, it follows that
\begin{eqnarray}
\left|\frac{\partial^2H}{\partial\theta^2(x)}\right|\le 2d\left(\sup_{\RR}\left|(g^i)''\right|+\frac{1}{\lambda}\right)\le 6d\bar{C}.
\end{eqnarray}
Hence
\begin{eqnarray}
\label{c13}
|A(k)|\le\frac{12d\bar{C}}{k^2}.
\end{eqnarray}
Using (\ref{c13}) for $|k|\ge (12d\bar{C})^{1/2}$ and (\ref{c11}) for $|k|\le (12d\bar{C})^{1/2}$, we get (\ref{c10}).

~

Finally, if $x=0$ we note that
\begin{eqnarray}
F_k(\theta)=-\frac{1}{k^2}\frac{\partial^2}{\partial\theta^2(e_i)}F_k(\theta)
\end{eqnarray}
and we proceed as before.
\endpf
\begin{rem}\normalfont
\label{3.2}
The proof shows that for $h=g''$ we can also use norms involving only lower derivatives of $g$. In particular, we have
\begin{eqnarray}
\label{c14}
|\la g''\ra_H|&\le&\frac{1}{2\pi}\int_{\RR}|\hat{g''}(k)||A(k)|\rmd k\nonumber\\
&\le&\frac{1}{2\pi}||\hat{g'}(k)||_{L^2(\RR)}\left(\int_{\RR}k^2|A(k)|^2\rmd k\right)^{1/2}\nonumber\\
&\le&\frac{1}{\sqrt{2\pi}}||g'||_{L^2(\RR)}\left(2\left(\frac{1}{3}+(12d\bar{C})^2\right)\right)^{1/2},
\end{eqnarray}
where we used (\ref{c11}) for $|k|\le 1$ and (\ref{c13}) for $|k|\ge 1$.
\end{rem}
\begin{rem}\normalfont
Note that our proofs can be very easily adapted to any decomposition of $\mu=\mu_1\ast\mu_2$, where $\mu_1$ and $\mu_2$ are Gaussian with 
covariances $C_1$ and $C_2$, such that $H_{u,\psi}(\theta):=G(u,\psi+\theta)+\frac{1}{2}(C^{-1}\theta,\theta)$ is uniformly convex. 
\end{rem}
\begin{rem}\normalfont
The procedure for the one-step decomposition can be iterated and the proofs can be adapted to the multi-scale decomposition; iterating the method would lower the temperature and weaken 
the conditions on the pertubation function $g$. However, our iteration procedure would not allow us to get results involving the low temperature case.
\end{rem}
\subsection*{Examples}
\begin{enumerate}
\item [(a)] $V(s)=s^2+a-\log(s^2+a),~~\mbox{where}~~0<a<1$. Then $C_1=C_2=2$, $C_0=\frac{2}{a}$, $||(g^i_0)''||_{L^1(\RR)}=2\sqrt{\frac{1}{a}}$ 
and $\beta\le\frac{a^2\pi^2}{6\times 16^2 d}$.

\begin{figure}[!h]
\input{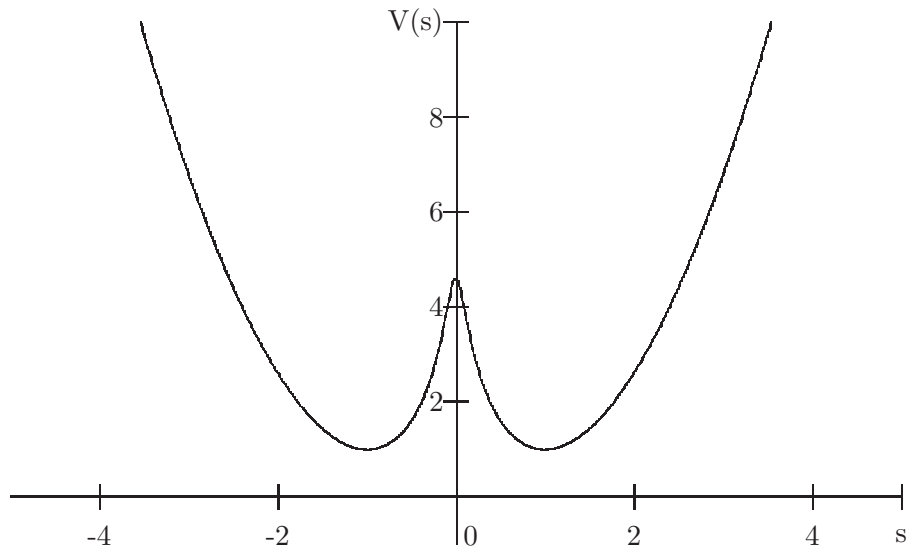}
\caption{Example (a)}
\end{figure}

\item [(b)] Let $0<\delta<1$ and

$$V(s)=
\left\{
\begin{array}{lcc}
\frac{x^2}{2}-\frac{4}{\delta^4}x^3(\delta-x)^3& \mbox{if} &0\le x\le\delta\\
\frac{x^2}{2} & \mbox{otherwise}.
\end{array}
\right.$$
Then $C_1=C_2=1$, $\bar{C}=\frac{6}{5}$, $||(g^i_0)''||_{L^1(\RR)}\le\frac{3}{10\sqrt{5}}\delta^5$ and $\beta\le\left(\frac{5\sqrt{5d}\pi}{2\delta}\right)^2$.

Note that if $\delta<<1$, the surface tension is convex for very large values of $\beta$.
\begin{figure}[!h]
\input{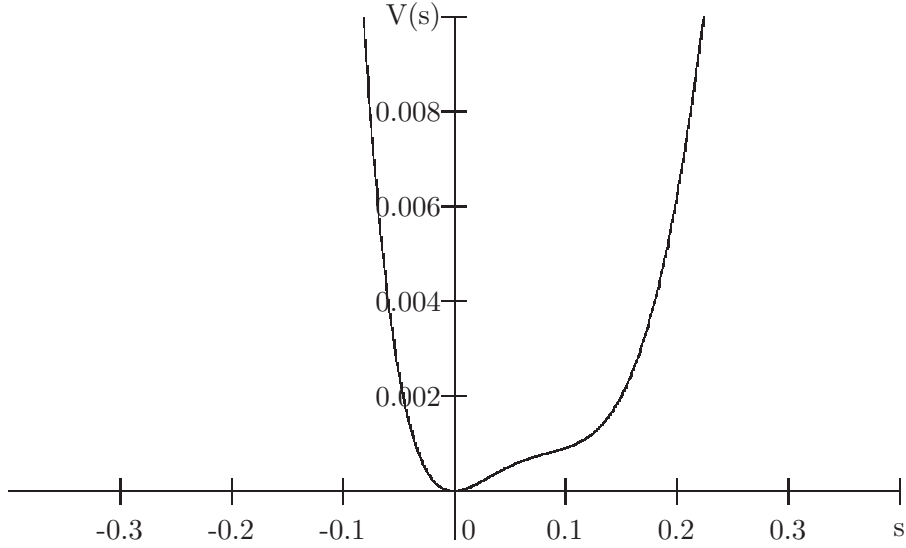}
\caption{Example (b)}
\end{figure}

\item [(c)] Let $p\in (0,1)$ and $0<k_2<k_1$. Let
\begin{eqnarray*}
V(s)=-\log\left(pe^{-k_1\frac{s^2}{2}}+(1-p)e^{-k_2\frac{s^2}{2}}\right).
\end{eqnarray*}
Then
\begin{eqnarray*}
V_0''(s)=\frac{pk_1e^{-k_1\frac{s^2}{2}}+(1-p)k_2e^{-k_2\frac{s^2}{2}}}{pe^{-k_1\frac{s^2}{2}}+(1-p)e^{-k_2\frac{s^2}{2}}}
\end{eqnarray*}
and
\begin{eqnarray*}
g_0''(s)=-\frac{p(1-p)(k_1-k_2)^2s^2}{p^2e^{-(k_1-k_2)\frac{s^2}{2}}+2p(1-p)+(1-p)^2 e^{(k_1-k_2)\frac{s^2}{2}}}.
\end{eqnarray*}
We have
\begin{eqnarray*}
k_2\le V_0''(s)\le pk_1+(1-p)k_2~~\mbox{and}~~-\frac{p(k_1-k_2)}{1-p}\le g_0''(s)\le 0,
\end{eqnarray*}
where the lower bound inequality for $g_0''(s)$ follows from the fact that $g_0''(s)$ attains its minimum for $s\ge\sqrt{\frac{2}{k_1-k_2}}$. Then
\begin{eqnarray*}
||g_0''(s)||_{L^1(\RR)}\le\frac{2p}{1-p}\sqrt{(k_1-k_2)\pi}~~\mbox{and}~~\beta\le\left(\frac{1-p}{16p}\right)^2\frac{\pi k_2}{12d\bar{C}(k_1-k_2)}.
\end{eqnarray*}
Note that example c) is the one used in \cite{BK} to prove that unicity of ergodic states can be violated for non-convex $V$ for large enough $\beta$.

\end{enumerate}

\section*{Acknowledgment}
Codina Cotar thanks David Brydges and Haru Pinson for invaluable advice and suggestions during the writing of the manuscript.

\end{document}